\documentclass{article}

\usepackage{arxiv}

\usepackage[utf8]{inputenc} 
\usepackage[T1]{fontenc}    
\usepackage{hyperref}       
\usepackage{url}            
\usepackage{booktabs}       
\usepackage{amsfonts}       
\usepackage{nicefrac}       
\usepackage{microtype}      
\usepackage{lipsum}
\usepackage{graphicx}

\usepackage{lineno}
\usepackage{caption}
\usepackage{subcaption}
\usepackage{braket}
\usepackage{cuted}
\usepackage{amsmath} 
\usepackage{mathrsfs}
\graphicspath{ {./images/} }

\title{Doppler-assisted quantum resonances through swappable excitation pathways in Potassium vapor}

\author{
 Gourab Pal \\
  Raman Research Institute\\
  Bangalore, India \\
  \texttt{gourab@rrimail.rri.res.in} \\
   \And
 Subhasish Dutta Gupta \\
  TIFR Hyderabad, IISER Kolkata\\
  IIT Jodhpur\\
  \texttt{sdghyderabad@gmail.com} \\
  \And
 Saptarishi Chaudhuri \\
 Raman Research Institute\\
  Bangalore, India \\
  \texttt{srishic@rri.res.in} \\
}

\begin{document}
\maketitle
\begin{abstract}
We report the observation of two additional sub-natural line width quantum interference in the $D_2$ manifold of $^{39}K$ vapor, in addition to the usual single Electromagnetically induced transparency peak. The other two features appear exclusively because $^{39}K$ ground hyperfine splitting is smaller than the Doppler broadened absorption profile. This allows probe and control beams to swap their transition pathways. The control beam detuning captures the nature of the coherence, therefore an unusual phenomenon of conversion from perfect transparency to enhanced absorption is observed and explained by utilizing adiabatic elimination of the excited state in the Master equation. Controlling such dark and bright resonances leads to new applications in quantum technologies viz. frequency offset laser stabilization and long-lived quantum memory.
\end{abstract}


\section{Introduction}
For the past few decades, manipulating and controlling an atomic medium's optical responses \cite{quantum_coherence_book} using quantum interference across excitation channels has been a versatile field of study. One of the most fascinating quantum interference phenomena is electromagnetically-induced transparency (EIT) \cite{Harris1997, marangose, Finkelstein_2023}, which modifies the medium's dispersion characteristics dramatically and opens the door to several cutting-edge uses in quantum information processing \cite{Yang2013}, including the creation of high-precision quantum sensors \cite{Kitching2018}, atomic clocks \cite{Vanier2005}, magnetometers \cite{Vanier2005}, etc. As an application to quantum memory \cite{Ma_2017}, it is feasible to slow down \cite{Jen:13} or store photons \cite{Zhang:11} for a considerable time due to the steep anomalous dispersion in an EIT medium. The EIT medium can be used for creating entangled bi-photons\cite{Shu2016} having application in quantum communication. More recently, exotic effects Goos-Hänchen \cite{Soni:14}, and Imbert-Fedorov shifts \cite{PhysRevA.91.033831} have been observed employing EIT medium.

\begin{figure}[!htb]
     \centering
     \begin{subfigure}[b]{0.5\textwidth}
         \centering
         \fbox{\includegraphics[width=\textwidth]{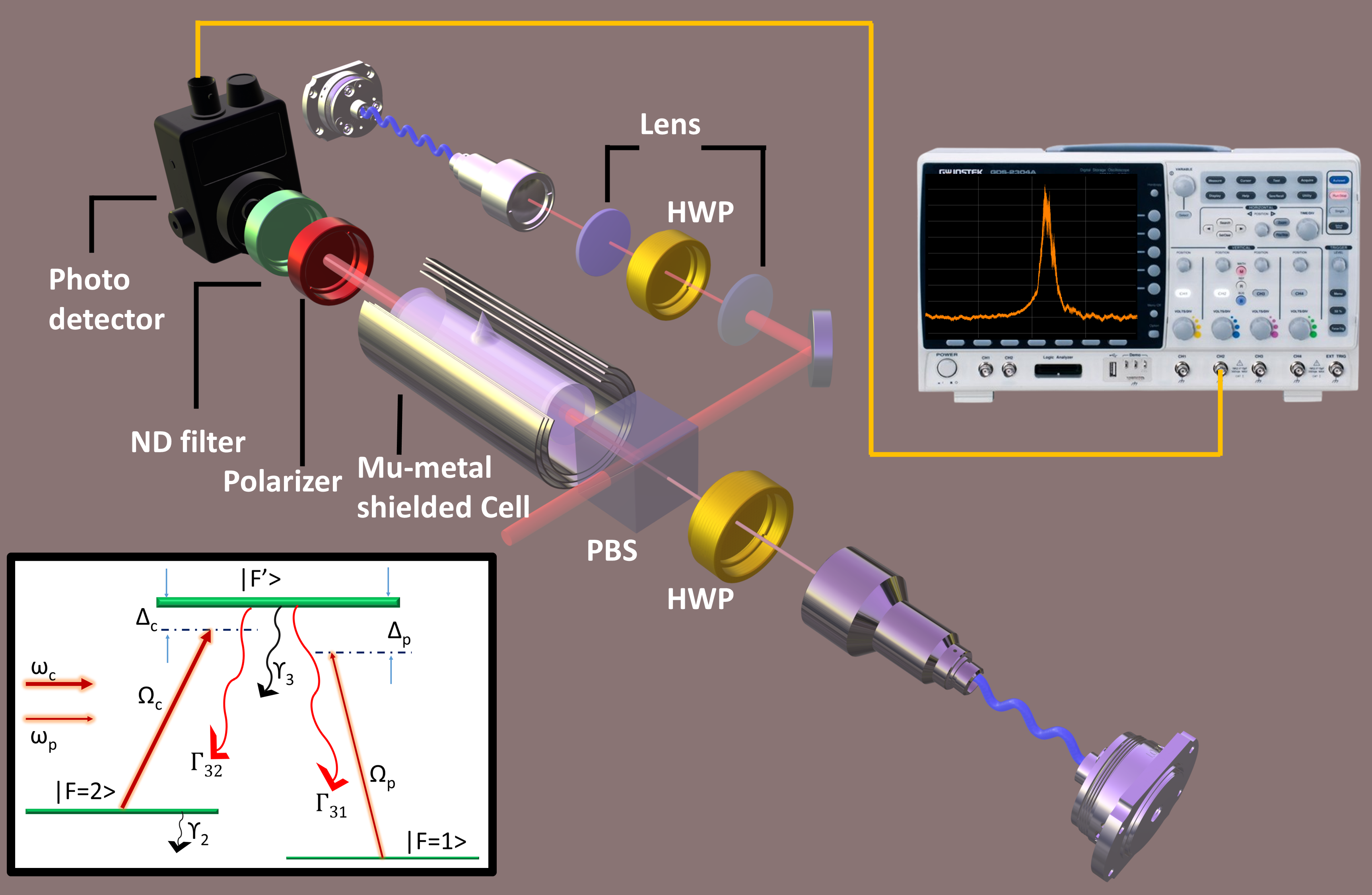}}
         
         \label{schematic}
     \end{subfigure}
     \hfill
     \begin{subfigure}[b]{0.5\textwidth}
         \centering
         \fbox{\includegraphics[width=\textwidth]{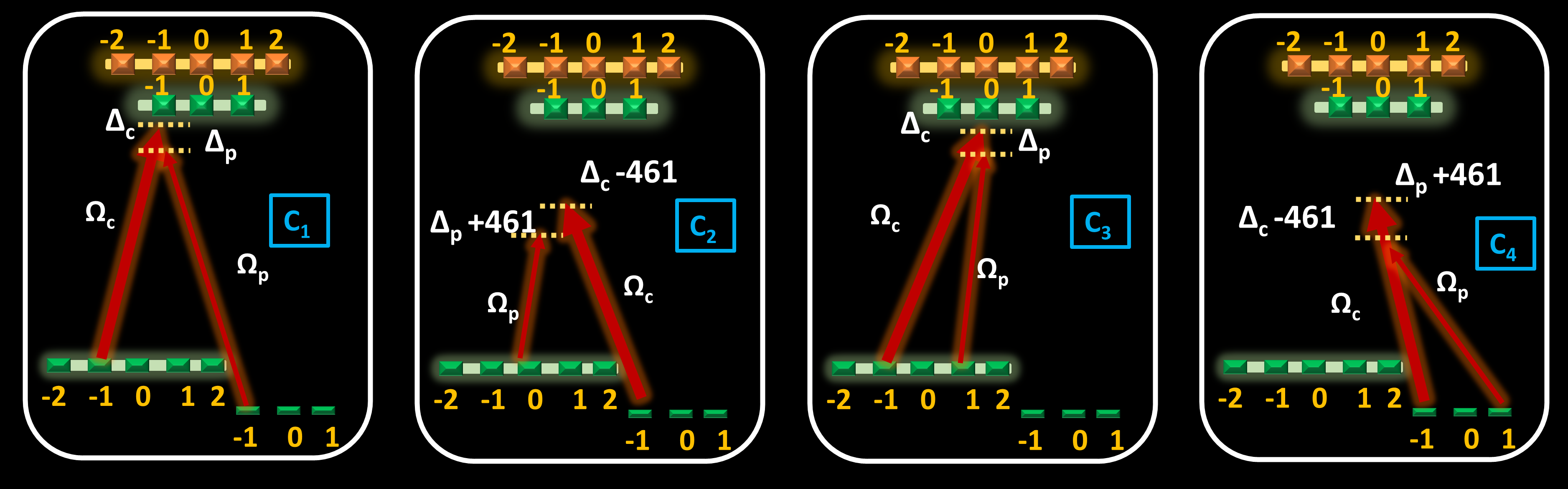}}
        
         \label{energy_levels}
     \end{subfigure}
        \caption{(Up) Schematic of experimental setup. Essential optical components are labeled in the figure. The inset shows a typical three levels $\Lambda$-system with relevant parameters. (Down) All four $\Lambda$-type configurations are supported due to small ground state splitting (see text) in $^{39}K$ atomic ensemble. Here ground states are $\ket{F=1}$ (indicated by a thin line) and $\ket{F=2}$ (indicated by a relatively thick line). (color online)}
        \label{f1}
\end{figure}

Numerous experiments have been carried out in diverse atomic systems such as $^{87}Rb$\cite{Thomas:13}, $^{133}Cs$ \cite{SARGSYAN20122090}, $^{23}Na$ \cite{PhysRevA.87.023836}, metastable He \cite{10.1117/12.2616988}, Erbium-doped fiber  \cite{Bencheikh:10}, molecular Lithium \cite{PhysRevA.82.023812}, EIT with Rydberg atoms \cite{Cheng:17, B.S.:22}. These systems exhibit strong quantum interference phenomena (like EIT) due to their wide ground state separation and considerable energy spacing among excited states. Furthermore, in a typical atomic vapor, the Doppler broadening is much smaller than the ground state separation, leading to the observation of just one EIT peak. On the other hand, the ground state separation in $^{39}K$ is smaller (461.7MHz \cite{Tiecke2011PropertiesOP}) than the Doppler broadening at room temperature which makes the study of EIT in $^{39}K$ intriguing. There exist very few quantum interference-based experiments in $^{39}K$, Long et.al. \cite{Long:17} used a chirped waveform electro-optic modulator to investigate the EIT, A. Sargsyan et al. \cite{Sargsyan2016} demonstrated on-resonance EIT and Gozzini et.al \cite{Gozzini:17} observed EIT-EIA transition in Hanle configuration with polarization as a tuning parameter. In this article, we experimentally demonstrate the emergence of three distinct quantum interference line shapes in a $\Lambda$-type system in $^{39}K$ atomic ensemble. The ground-state hyperfine splitting being smaller than the unresolved two overlapping Doppler absorption profiles, the excitation pathways of the control and probe beam can be swapped. We also observe a transition from complete transparency to enhanced absorption as we vary the control beam detuning from the blue side to the red side of its resonance point, while the two-photon detuning condition is satisfied.

\begin{figure}[ht]
\centering
\fbox{\includegraphics[width=0.5\textwidth]{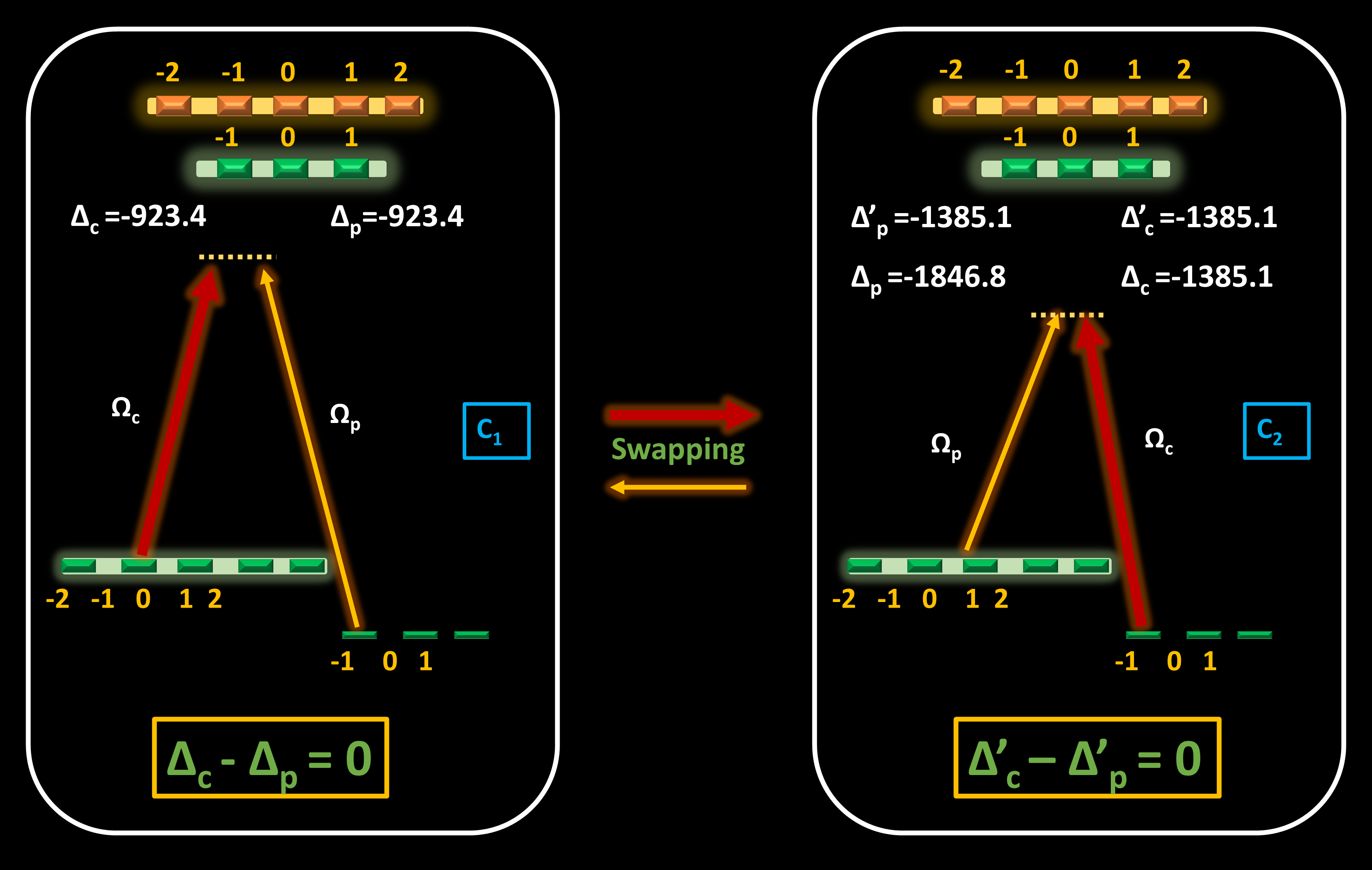}}
\caption{(Left) The usual case $C_1$ with two-photon detuning $\Delta_c-\Delta_p=0$. (Right) Control and probe transition pathways are swapped. In the swapped frame (case $C_2$), $\Delta^{'}_c-\Delta^{'}_p=0$. See text for details. (color online)}
\label{f2}
\end{figure}

\section{Experimental setup and methods}

Figure \ref{f1} (Upper panel) shows a schematic of the experiment. Two independent grating stabilized diode lasers (Toptica DL Pro and DL100) are used as the control and probe beam. The probe and control lasers are tuned near F=1 to F' and F=2 to F' respectively in the $D_2$ manifold of $^{39}K$, with a typical laser line width of 1MHz. Two separate acousto-optic modulators are used to independently control the intensity and detuning of the probe and control beams. The frequency reference is obtained using two saturated absorption spectroscopy, see supplementary \cite{supplementary}. A 75 mm long and 25mm diameter spectroscopy-grade commercial vapor cell containing $^{39}K$ (in natural abundance, $\sim 93\%$) is placed inside a three-layered $\mu$-metal jacket to efficiently shield any stray magnetic field. The probe beam and control beams are mode-filtered using single mode polarization maintaining fibers and passed through two independent telescopic arrangements with exit beam diameters of 1mm and 4mm, respectively. The two beams are prepared in orthogonal linear polarization states and merged using a polarizing beam splitter cube before being sent to the atoms. The probe laser is kept in frequency scan mode whereas the control laser frequency is kept at various detunings. In the end, the control beam is filtered out using a linear polarizer (extinction ratio $1:10^5$), and the signal is obtained using a photo-detector (Thorlabs PDA10A2) with 150MHz bandwidth. A high-speed oscilloscope (2Gs/sec) is used for visualization and data recording. 

\begin{figure}[ht]
     \centering
     \begin{subfigure}[b]{0.5\textwidth}
         \centering
         \fbox{\includegraphics[width=\textwidth]{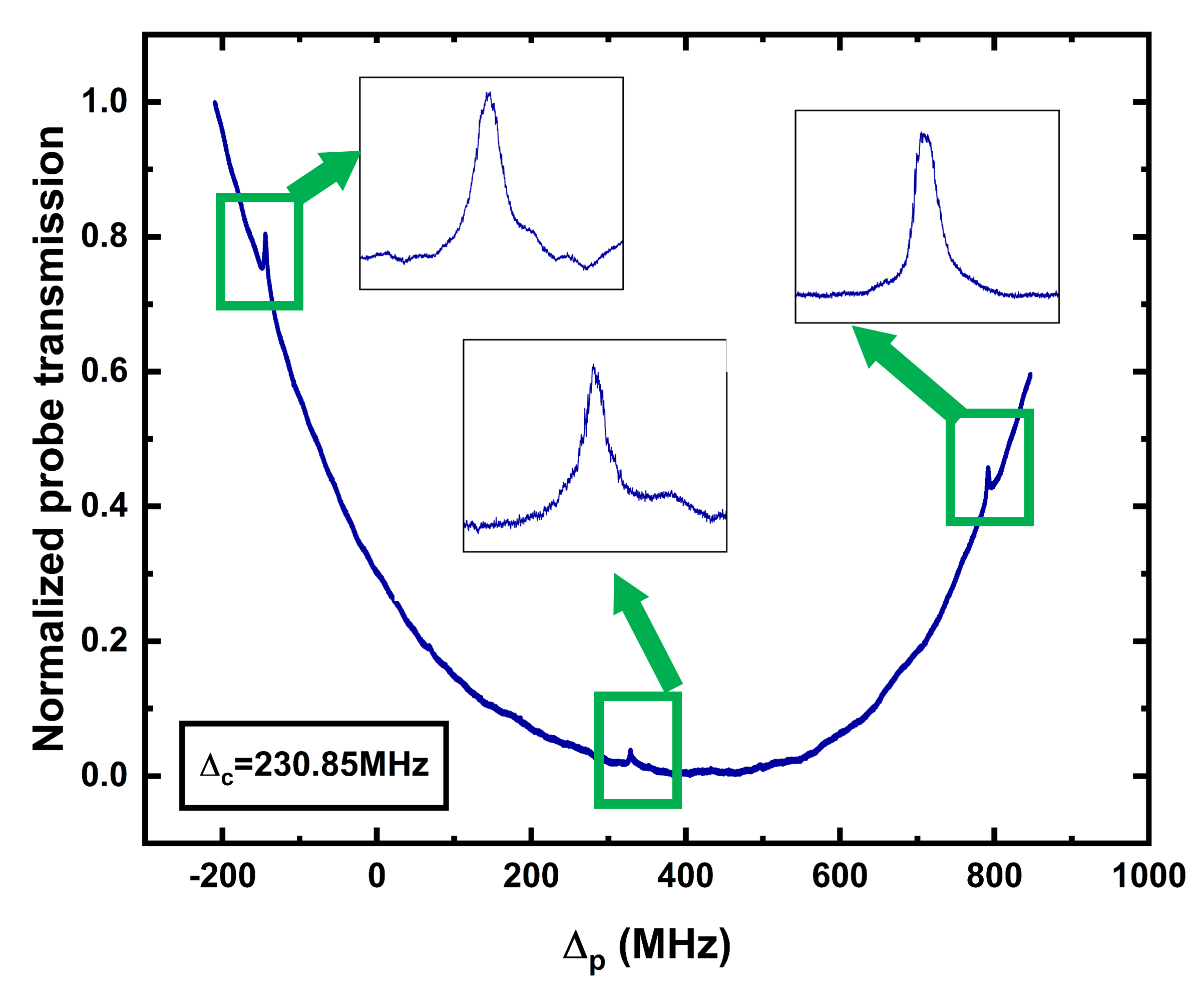}}
         
         \label{three}
     \end{subfigure}
     \hfill
     \begin{subfigure}[b]{0.5\textwidth}
         \centering
         \fbox{\includegraphics[width=\textwidth]{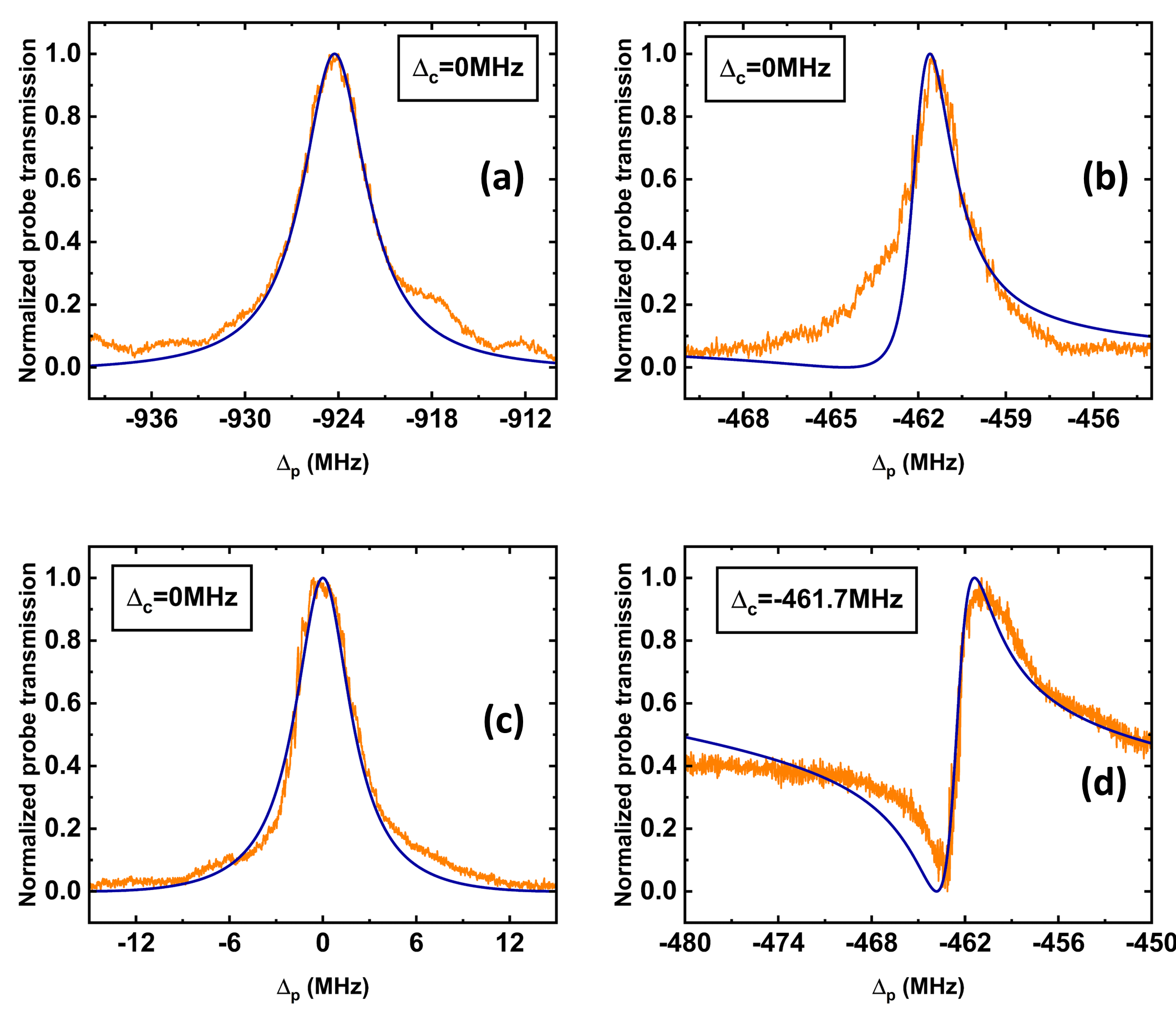}}
        
         \label{on}
     \end{subfigure}
        \caption{(Upper panel) Observation of three bright resonances for $\Delta_c=230.85$ MHz. The inset shows the zoomed views. (Lower panel) On-resonant ($\Delta_c=0$) bright resonance line shapes (a,b,c). The orange line indicates experimental data and the blue line is the theoretical model fit. Subplot (d) shows the admixture of dark and bright resonance at $\Delta_p=-461.7$ MHz. (color online)}
        \label{f3}
\end{figure}

\section{Theoritical description}
A semi-classical approach with density matrix formalism is applied to this three-level $\Lambda$ system as shown in the inset of figure \ref{f1} (Upper panel). The two ground states are $\ket{F=1}$ and $\ket{F=2}$ hyperfine levels in $^2S_{1/2}$ and the excited state is taken as unresolved $\ket{F'}$ in $^2P_{3/2}$ containing F'=0,1,2 and 3 states. The Master equation that governs the dynamics of a mixed state is, $\dot{\rho} = - \frac{i}{\hbar} [H,\rho]+ \mathcal{L} \rho$. Here $\rho$ is the density matrix of the system, and H is the EIT Hamiltonian \cite{marangose}. The last term $\mathcal{L} \rho$, is the pertinent Lindbladian term that controls all potential decay processes such as leaky transitions from other excited states, spontaneous emission, and decoherence resulting from Doppler-broadening. Such decay terms account for the interactions with the environment and thus make the system's dynamics non-unitary. In our case, the dominant decay processes are due to Doppler broadened spontaneous emission and energy-conserving dephasing effects like collision of atoms with walls, and atom-atom collisions. 

The non-unitary Lindbladian term, see supplementary \cite{supplementary}, is $\mathcal{L} \rho = \Gamma_{31} \mathcal{D}[\hat{\sigma}_{1,3}]\rho + \Gamma_{32} \mathcal{D}[\hat{\sigma}_{2,3}]\rho + \gamma_{2} \mathcal{D}[\hat{\sigma}_{2,2}]\rho + \gamma_{3} \mathcal{D}[\hat{\sigma}_{3,3}]\rho$,  where $\Gamma_{ij}$ denotes spontaneous decay rate from state i to state j, $\gamma_{i}$ is a energy-conserving dephasing term for state i, $\hat{\sigma}_{i,j}$ is a jump operator defined as $\hat{\sigma}_{i,j} = \ket{i}\bra{j}$ and 
 $\mathcal{D}$ is coined as Lindblad super-operator whose action is defined as $\mathcal{D}[A]B=ABA^\dag - \frac{1}{2} \{ A^\dag A,B\}$ for any two operators A and B. In light of such formalism, the susceptibility seen  by the probe $\chi(\omega_p)$ can be obtained by solving coherence terms under steady-state conditions as,

\begin{equation} 
\label{probe susceptibility}
        \chi(\omega_p) = \frac{2n|d_{13}|}{\epsilon_0 E_p} \frac{\Omega_p[2i\Delta+\gamma_2]}{-4i\Delta_p\Delta+i\Omega_c^2-2\Delta_p\gamma_2+i\Gamma_3[2i\Delta+\gamma_2]}
\end{equation}

\begin{figure}[htb!]
\centering
\fbox{\includegraphics[width=0.45\textwidth]{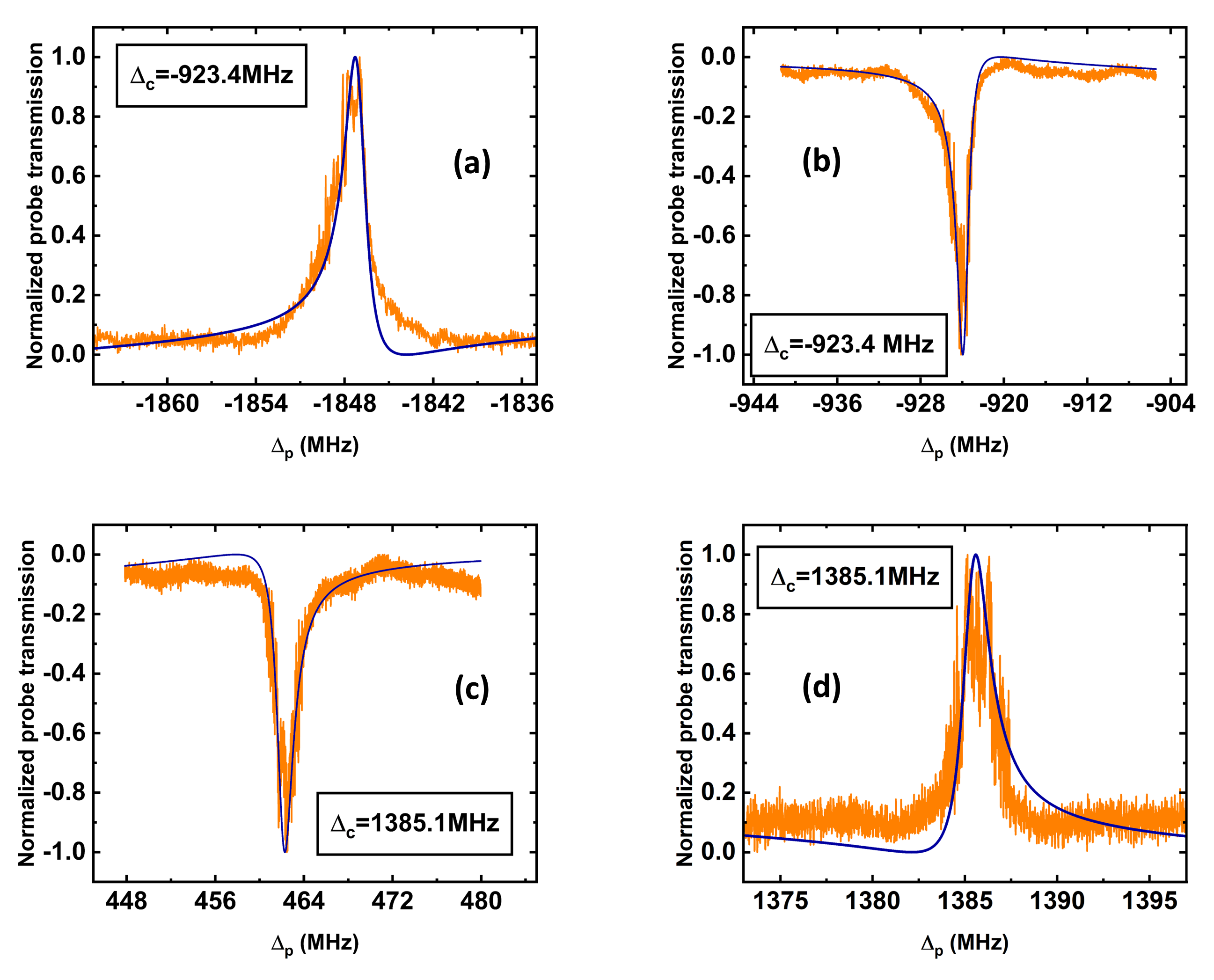}}
\caption{(Upper panel) In the red-side of control beam resonance, subplot a shows complete transparency due to case $C_2$, subplot b shows full absorption due to case $C_1$. (Lower panel) The opposite parity line shapes are seen with blue-detuned control light as shown in subplot c and d.}
\label{f4}
\end{figure}

Here,  n is the density of atoms at cell temperature T, $|d_{13}|$ is the dipole matrix element, $E_p$ is probe field amplitude,$\Gamma_3=\Gamma_{32}+\Gamma_{31}+\gamma_3$ is the total decoherence of excited state, $\gamma_{2}$ indicates ground state decoherence, $\gamma_{3}$ is excited state energy conserving decoherence, $\Omega_p$, $\Omega_c$ are probe and control Rabi flopping frequencies, respectively and $\Delta_p$, $\Delta_c$ and $\Delta=\Delta_p-\Delta_c$, are respective probe detuning, control detuning and two-photon detuning from a given transition (see figure \ref{f1}). 

\begin{figure}[htb!]
\centering
\fbox{\includegraphics[width=0.5\textwidth]{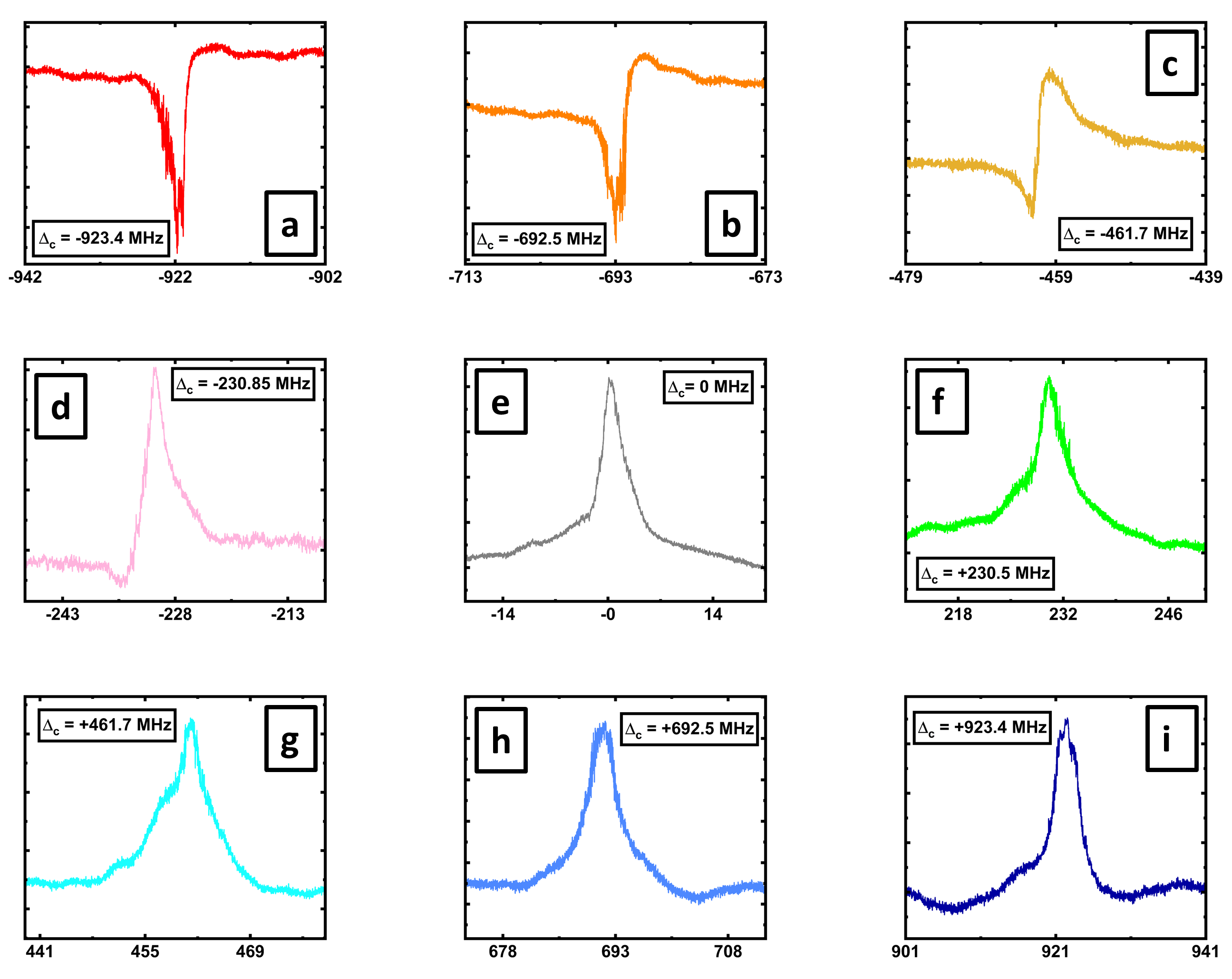}}
\caption{Transition from bright resonance to dark resonance as we vary the control detuning for the case $C_1$. The x-axis is the probe detuning, y-axis represents probe transmission. Inset-box shows the corresponding $\Delta_c$ values. (color online)}
\label{f5}
\end{figure}

\section{Results and discussions}
The ground state splitting of the Potassium atom is 461.7 MHz which is smaller than the unresolved two Doppler broadened absorption profiles (684 MHz \cite{supplementary} at a cell temperature of $60^o$C). Under this circumstance, the usual EIT peak with $\Delta_c=0$ is observed which is the standard case $C_1$ in figure \ref{f1}. However,
since there exists a significant number of atoms with velocities such that a Doppler shift can be higher than ground-state hyperfine splitting, the excitation pathways of the control and probe beam can now be swapped. The swapping mechanism is illustrated in figure \ref{f2} where, as an example, $\Delta_c=-923.4$MHz is taken. Before swapping, $\Delta_c=\Delta_p=-923.4$ MHz, so the two-photon detuning is satisfied. However after swapping, $\Delta^{'}_c=\Delta_c-461.7=-1385.1$ and $\Delta^{'}_p=\Delta_p+461.7$. Since the probe laser is scanning, the probe frequency is adjusted to $\Delta_p=-1846.8$ so that $\Delta^{'}_p=-1385.1$ and the two-photon detuning condition is satisfied in this swapped frame. A third line shape is obtained by applying the same arguments but with control and probe beam making a valid $\Lambda$ from the same hyperfine state either $\ket{2}$ (for the case $C_3$) or $\ket{1}$ (for the case $C_4)$ \cite{supplementary}. See figure \ref{f1} (lower panel) for all four possible configurations. There are multiple $\Lambda$-system within a given case, considering all possible magnetic sublevels. We have considered all such transition probabilities with their corresponding Clebs-Gordon coefficients. The resulting line shapes and their positions remain largely unaffected as discussed in \cite{supplementary}.

 As a representative dataset, figure \ref{f3} (Upper panel) shows all three EIT line shapes riding on the Doppler broadened absorption profile, here the control detuning is set to $\Delta_c=230.85$ MHz. The corresponding background-subtracted features are zoomed in the insets. The relevant physics is well captured by introducing correct dephasing terms. The figures (a,b,c) in figure \ref{f3} (lower panel) show all possible bright resonances with corresponding theoretical model fit. The subplot (c) is the usual single EIT peak, which is also observed in other atoms like Rb, Cs, etc. The characteristic variation of EIT line width with $\Omega_c$ for the case of $\Delta_c=-923.4$ MHz is also studied with theoretical fitting (see supplementary \cite{supplementary}) as presented in the susceptibility expression. The inclusion of an overlapping Doppler broadened profile enables us to exchange the excitation pathways of control and probe which give two additional features (a and b). The asymmetric mismatch between theory and experiment is accounted for non-zero probe detuning and the existence of multiple excited states \cite{PhysRevA.83.053809, Chen2013} in the case of $^{39}K$. The asymmetry primarily originates from the admixture of pure dark state and pure bright state as can be seen in subplot (d) of figure \ref{f2} where the dip is observed due to bright resonance and the peak is due to dark resonance.

When the control beam detuning is taken to far-off red detuned, $\Delta_c=-923.4$MHz, a complete absorption and complete transmission line shapes are observed at probe detunings $\Delta_p=-923.4$MHz and $\Delta_p=-1846.8$MHz respectively. For blue-detuning, $\Delta_c=1385.1$MHz, complete absorption, and transmission are seen in $\Delta_p=461.7$MHz and $\Delta_p=1385.1$MHz respectively as shown in figure \ref{f4}. For the cases of far-off resonant complete absorption, we have adiabatically eliminated \cite{Brion_2007} the state $\ket{F'}$ from the Master equation to get the correct line shape by switching off the spontaneous decay in the Lindblad term. This is a reasonable description because, at far-off resonance from the one-photon transition the lasers do not significantly populate the excited state thereby reducing the spontaneous decay from the excited state. The theoretical model fit gives effective $\gamma_2$ values ranging from 1.2 MHz to 2.5 MHz depending on the line shape position. The other variable parameters $\gamma_3, \Gamma_{31}, \Gamma_{32}$ are adjusted according to the probe and control detunings. For a given line shape, if it appears inside the Doppler, then the dephasing will be controlled by the spontaneous decay term and if it appears outside the Doppler, then the dephasing will come primarily from $\gamma_{3}$. As an example, for $\Delta_c=-923.4$ MHz (case $C_3$), the values of $\gamma_2=1.2$ MHz, $\Gamma_{31}=\Gamma_{32}=0$ and $\gamma_3=342$ MHz.

\begin{figure}[htb!]
\centering
\fbox{\includegraphics[width=0.5\textwidth]{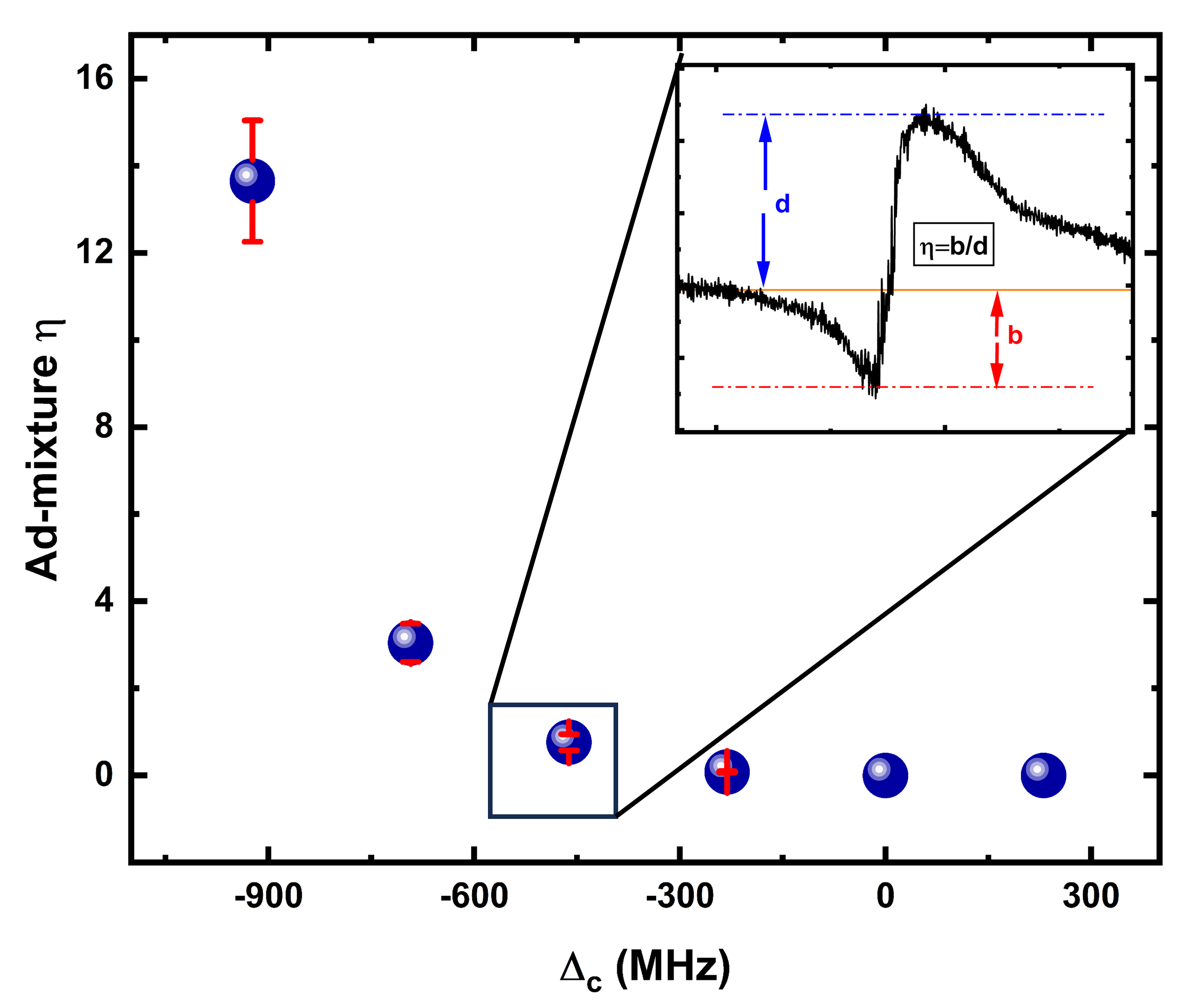}}
\caption{Ad-mixture parameter $\eta$ as a function of control detuning $\Delta_c$. As $\Delta_c$ is swept from the red side to the blue side, complete transparency is obtained due to the formation of dark resonance, consequently, the bright state ad-mixture parameter $\eta$ vanishes after $\Delta_c \geq 0$.}
\label{f6}
\end{figure}

Figure \ref{f5} (a-i) shows a transition from bright resonance to dark resonance as we vary the control beam detuning keeping the probe laser scanning across the whole $D_2$ manifold. The blue side of this case always gives bright resonance as can be seen from (f-i) of figure \ref{f5}. The line-plot colors are chosen to represent the detuning of the control beam. These line shapes are obtained for the case $C_1$. A similar transition can be found in other cases also. 
To indicate the extent of mixing of bright and dark resonance, the ad-mixing parameter $\eta$ is defined as the ratio of bright resonance contrast (defined as b) and dark resonance contrast (defined as d). The variation of $\eta$ is plotted against the control beam detuning which gave exponential crossover from bright resonance to dark resonance as we sweep the control beam detuning as shown in figure \ref{f6}.

\section{Conclusion and outlook}

By choosing both the Doppler and exchanging the probe and control beam excitation routes, we have been able to explain all possible line shapes and positions and they match well with the density matrix formalism-based theoretical model. The adiabatic elimination works well to describe the bright resonance in blue as well as red detuning regions. The exact line shape can be matched by taking into account the intervention of other excited states as described in \cite{Subba2020}. Going beyond the Gaussian probe and control, a probe beam carrying orbital angular momentum (Laguerre Gaussian mode) can have a fundamentally unique response \cite{Wang:22} to this EIT medium because of the spatial variation of Rabi frequency that can lead to line shape narrowing \cite{PhysRevA.81.043804} as well as change in asymmetries. Another novel avenue to explore EIT in Potassium will be to study cold Potassium atoms \cite{dual_species_mot} in a strong magnetic field where the external magnetic field can couple the Zeeman sub-levels of different hyperfine states. The explanation of all these quantum interference line shapes enriches the understanding of quantum coherence and has applications to frequency-offset tight laser locking \cite{Ying:14} and quantum technology applications such as optical isolators \cite{Weller:12} quantum memory \cite{Ma_2017}, slow light \cite{Khurgin:10}.

\medskip
\medskip
\medskip

{\huge\bfseries Appendices}
\medskip

\appendix

\section{The Master equation}

In the density matrix formalism, the Von Neumann equation governs the time evolution of mixed-state density matrix $\rho$
\begin{equation}
    \dot{\rho} = -\frac{i}{\hbar} [H,\rho]
\end{equation}
Here H is the EIT Hamiltonian \cite{marangose} for our 3-level $\Lambda$ system as described in the main text.

\begin{equation} \label{eit hamiltonian}
    H=-\frac{i}{\hbar}\begin{bmatrix}
        0 & 0 & \Omega_p\\
        0 & 2(\Delta_p-\Delta_c) & \Omega_c\\
        \Omega_p & \Omega_c & 2\Delta_p
    \end{bmatrix}
\end{equation}

In atomic systems, there are many decay processes such as spontaneous emission, dephasing, or decoherence due to inhomogeneous magnetic fields, ground state collisions, spin-exchange collisions, and so on. Such terms make the above Hamiltonian non-unitary. We can make a Lindblad term that needs to be added to the above Hamiltonian to correctly describe our present system.
The irreversible Lindblad term can be written as 

\begin{equation}
    \mathcal{L} \rho = \Gamma_{31} \mathcal{D}[\hat{\sigma}_{1,3}]\rho + \Gamma_{32} \mathcal{D}[\hat{\sigma}_{2,3}]\rho + \gamma_{2} \mathcal{D}[\hat{\sigma}_{2,2}]\rho + \gamma_{3} \mathcal{D}[\hat{\sigma}_{3,3}]\rho
\end{equation}

\begin{figure}[ht]
\centering
\fbox{\includegraphics[width=.5\linewidth]{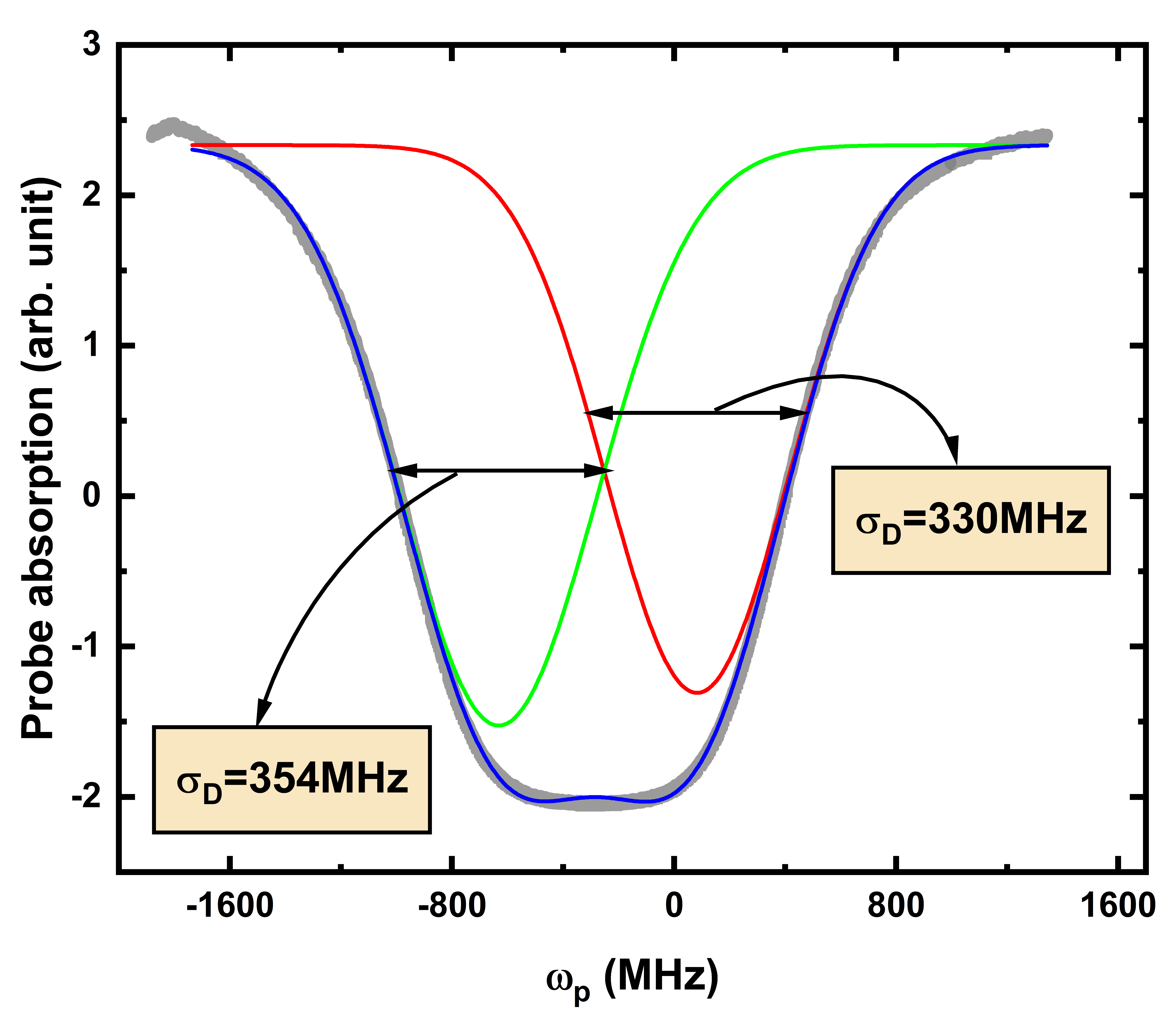}}
\caption{Representation of overlapping Doppler broadened absorption profiles. The thick grey line is the experimental data, the red and green are individual Doppler absorption fitted by Gaussian, and the blue line is the convolution of two Doppler. (color online)}
\label{two_dop}
\end{figure}

where $\Gamma_{ij}$ denotes spontaneous decay from state i to state j, $\gamma_{i}$ is a energy-conserving dephasing term for state i, $\hat{\sigma}_{i,j}$ is a jump operator defined as $\hat{\sigma}_{i,j} = \ket{i}\bra{j}$ and 
 $\mathcal{D}$ is coined as Lindblad super-operator whose action is defined as $\mathcal{D}[A]B=ABA^\dag - \frac{1}{2} \{ A^\dag A,B\}$ for any two operators A and B. Setting the basis as $\{[1,0,0]^T, [0,1,0]^T, [0,0,1]^T \}$, the jump operators can be easily constructed. Upon simplification, the Lindblad term becomes

 \begin{equation} \label{lindblad}
 \mathcal{L} \rho=-\frac{1}{2}
     \begin{bmatrix}
         0 & \gamma_2 \rho_{12} & (\Gamma_{31}+\Gamma_{32}+\gamma_3)\rho_{13}\\
         \gamma_2 \rho_{12}^* & 0 & (\Gamma_{31}+\Gamma_{32}+\gamma_3+\gamma_2)\rho_{23}\\
         (\Gamma_{31}+\Gamma_{32}+\gamma_3)\rho_{13}^* & (\Gamma_{31}+\Gamma_{32}+\gamma_3+\gamma_2)\rho_{23}^* & 0
     \end{bmatrix}
 \end{equation}
Here in $\rho$ matrix, we need to set $\rho_{11}=1$, $\rho_{22}=0$ and $\rho_{33}=0$.

\begin{figure}[htbp]
\centering
\fbox{\includegraphics[width=.5\linewidth]{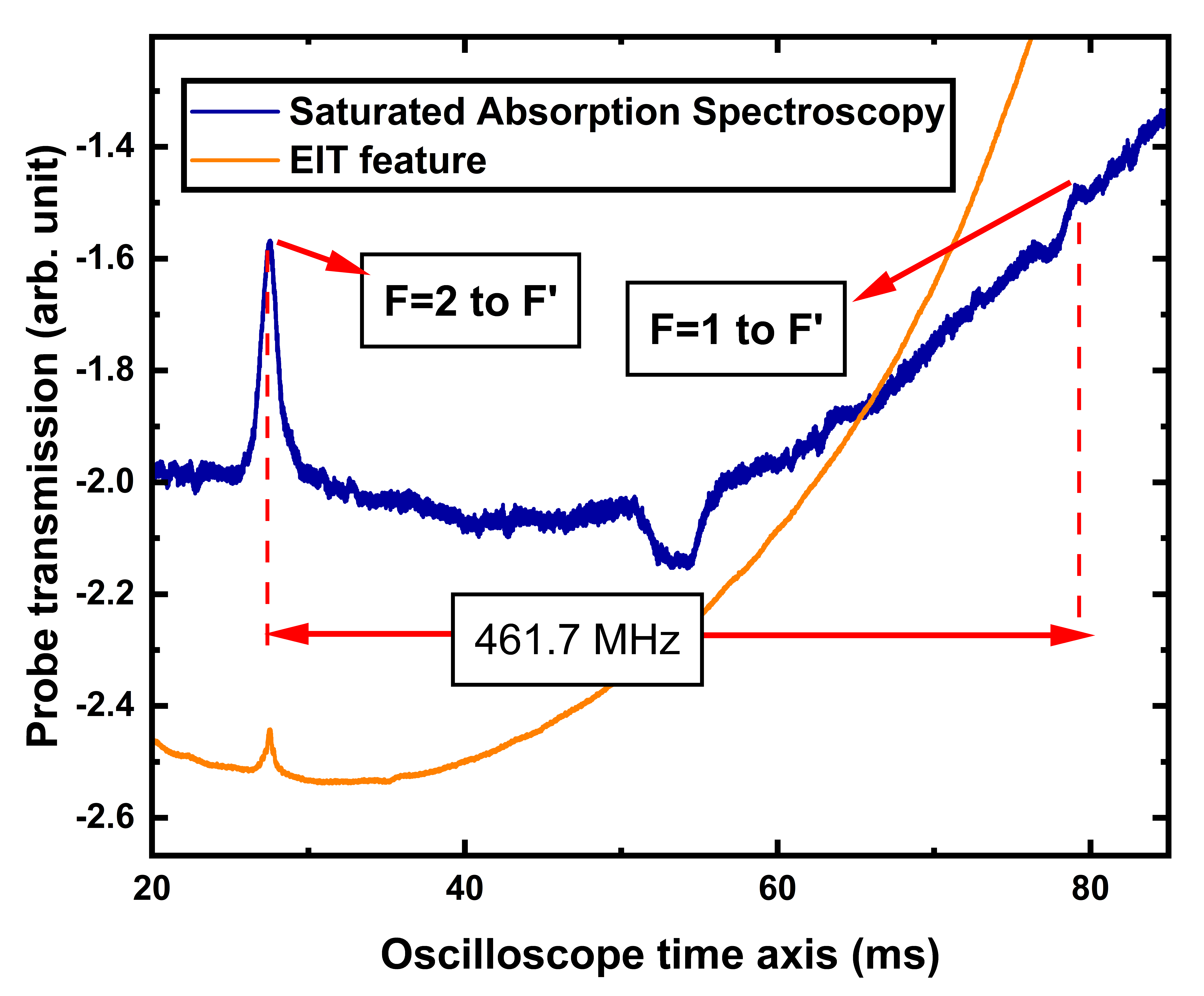}}
\caption{Saturated Absorption Spectroscopy of $^{39}K$ used for the conversion of time scales in oscilloscope to frequency scale.}
\label{sas}
\end{figure}

\section{Steady state solution of coherence terms}
After adding Lindbladian to the Von Neumann equation, we can read out two required coherence terms
\begin{equation}
    \dot{\rho}_{12} = [-\frac{1}{2}\gamma_2+i(\Delta_c-\Delta_p)]\rho_{12}-\frac{1}{2}i\Omega_c \rho_{13}
\end{equation}

\begin{equation}
    \dot{\rho}_{13} = [-\frac{1}{2}(\Gamma_{31}+\Gamma_{32}+\gamma_3)-i\Delta_p]\rho_{13}-\frac{1}{2}i\Omega_c \rho_{12}-\frac{1}{2}i\Omega_p
\end{equation}

In steady state conditions, we can set $\dot{\rho}_{ij}=0$, which gives two linear equations that can be solved analytically.
The solution for $\rho_{13}$ is directly related to the susceptibility seen by the probe.

\begin{equation} \label{probe_sus}
    \chi_p \sim \frac{\Omega_p[2i(\Delta_p-\Delta_c)+\gamma_2]}{-4i\Delta_p(\Delta_p-\Delta_c)+i\Omega_c^2-2\Delta_p\gamma_2+i(\Gamma_{31}+\Gamma_{32}+\gamma_3)[2i(\Delta_p-\Delta_c)+\gamma_2]}
\end{equation}
In the main text, we only plot the probe transmission in various detunings which is the imaginary part of $\chi(p)$.

\section{Overlapping Doppler broadened absorption profile in $^{39}K$}
The ground state hyperfine splitting is 461.7MHz. Now, at 333K, the Doppler width is 342 MHz, calculated using the formula $\sigma_D=\sqrt{\frac{k_BT}{mc^2}}\omega_p$, where $k_B$ is Boltzmann constant, m is atomic mass, c is the velocity of light and $\omega_p$ is the probe frequency in Hz.

Figure \ref{two_dop} shows the convolution of two Doppler broadened absorptions for $\ket{F=1}\rightarrow\ket{F'}$ and $\ket{F=2}\rightarrow\ket{F'}$. For theoretical modeling, we have taken the Doppler decoherence $\gamma_3=342$ along with total spontaneous decay of excited state $\Gamma_3=\Gamma_{32}+\Gamma_{31}=684$ MHz when dealing with line shapes inside any of the Doppler. 

\begin{figure}[htbp]
\centering
\fbox{\includegraphics[width=.95\linewidth]{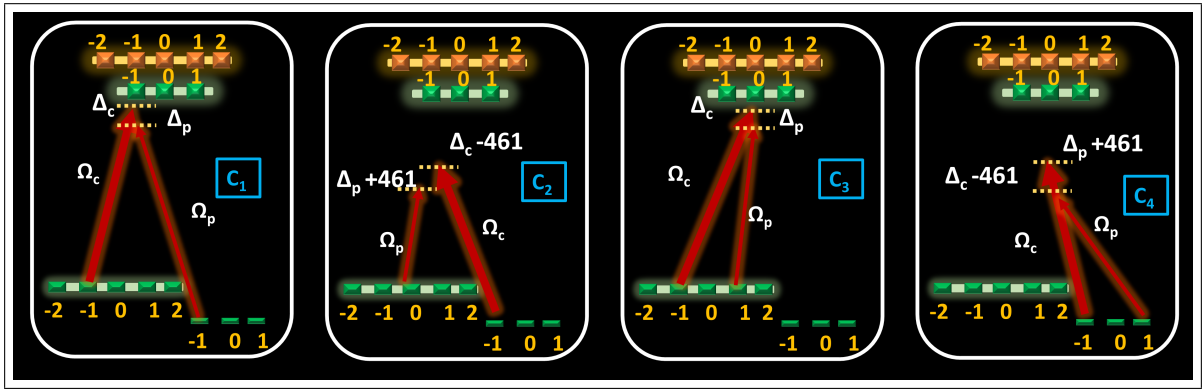}}
\caption{All four possible probe-control excitation configurations.}
\label{all_config}
\end{figure}

\section{Setting frequency reference}
Doppler-free Saturated Absorption Spectroscopy \cite{Mudarikwa_2012} is performed using an independent vapor cell. The blue line in figure \ref{sas} shows the hyperfine transition lines where the strongest peak corresponds to $\ket{F=2} \rightarrow \ket{F'}$ transition, the weakest peak corresponds to $\ket{F=1} \rightarrow \ket{F'}$ transition and the middle dip is the crossover resonance in the spectroscopy of $^{39}K$. The orange line shows a typical dark resonance. The separation between the two peaks in the spectroscopy refers to the ground hyperfine splitting as

\begin{equation}
    |T_{\ket{F=2}\rightarrow \ket{F'}}-T_{\ket{F=1}\rightarrow \ket{F'}}| = 461.7 MHz
\end{equation}
From the above equation, all time scales from the oscilloscopes can be converted to frequency scale.

\begin{figure}[ht]
\centering
\fbox{\includegraphics[width=.5\linewidth]{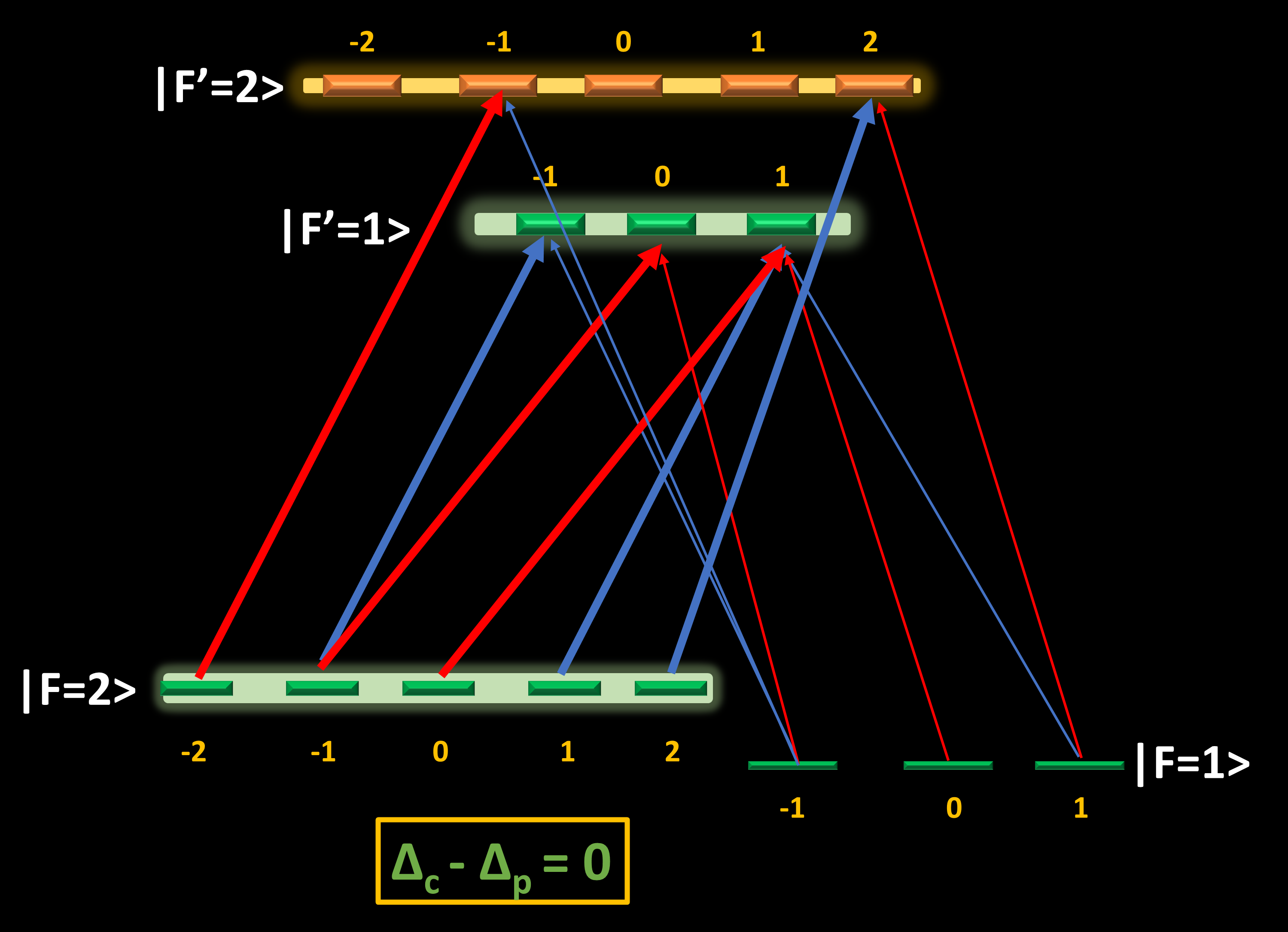}}
\caption{Demonstration of multiple $\Lambda$ systems for two types of polarization of light. Blue lines are used for $\pi$ polarized light and red lines are used for $\sigma$ polarized light. Only a few $\Lambda's$ are shown here for clarity. (color online)}
\label{magnetic}
\end{figure}

\section{Three line shapes}
Lets denote the probe susceptibility in equation \ref{probe_sus} as $\chi(\Delta_p, \Delta_c)$. Case $C_1$ is the standard case. For case $C_2$, we can exchange the role of probe and control in the following way 
$\chi(\Delta_p, \Delta_c)\rightarrow\chi(\Delta_p+461.7, \Delta_c-461.7)$. This new configuration still satisfies the two-photon resonance condition. In the same argument, the susceptibility for case $C_3$ is $\chi(\Delta_p+461.7, \Delta_c)$ and for case $C_4$ is $\chi(\Delta_p, \Delta_c-461.7)$. Finally, the net absorption can be written as 

\begin{equation}
    \alpha_p= \Sigma_{i=1}^4 \sum_{i=1}^k x_i \int_{v=-\infty}^{+v=\infty} Im(\chi)_i \mathcal{P}(v) \,dv 
\end{equation}
Here i is for case $C_i$ and $\mathcal{P}$ is the 1D Maxwell-Boltzmann velocity distribution given by
\begin{equation}
    \mathcal{P}=\frac{1}{\sqrt{2\pi} D} e^{-\frac{v^2}{D^2}}
\end{equation}
where $D=\sqrt{\frac{2k_BT}{m}}$ is the most probable velocity. Figure \ref{all_config} depicts all four cases.

\section{Contribution of all possible $\Lambda$-system due to magnetic sub levels}
Apart from having four different case as shown in figure \ref{all_config}, there are multiple $\Lambda$ systems possible within a given case due to the existence of near-degenerate magnetic sub-levels of each hyperfine states. Considering the configuration of case $C_1$, few $\Lambda$'s are shown in figure \ref{magnetic} where we use blue lines for linear polarized light and red lines for circular polarized light. Since we use linear polarization in our experiment, and a linear polarization state is a combination of right and left circular polarizations, all types of probe and control polarization choices need to be taken into account such as $Probe(\pi)-Control(\pi), Probe(\sigma^+)-Control(\sigma^+), Probe(\sigma^-)-Control(\sigma^-), Probe(\pi)-Control(\sigma^+), Probe(\pi)-Control(\sigma^-), Probe(\sigma^+)-Control(\pi), Probe(\sigma^-)-Control(\pi), Probe(\sigma^+)-Control(\sigma^-)$ and $Probe(\sigma^-)-Control(\sigma^+)$. The Rabi frequencies ($\Omega_p, \Omega_c$) are scales as $\Omega_c \rightarrow s\Omega_c$ and $\Omega_p \rightarrow s'\Omega_p$ when adding contributions where s and s' are the Clebs Gordon co-efficient \cite{Rb87_data} for the respective transitions. However, there is no significant difference in line shape and positions after considering all the possible $\Lambda$'s within a given case.

\begin{figure}[!htb]
     \centering
     \begin{subfigure}[b]{0.45\textwidth}
         \centering
         \fbox{\includegraphics[width=\textwidth, height=4.5cm]{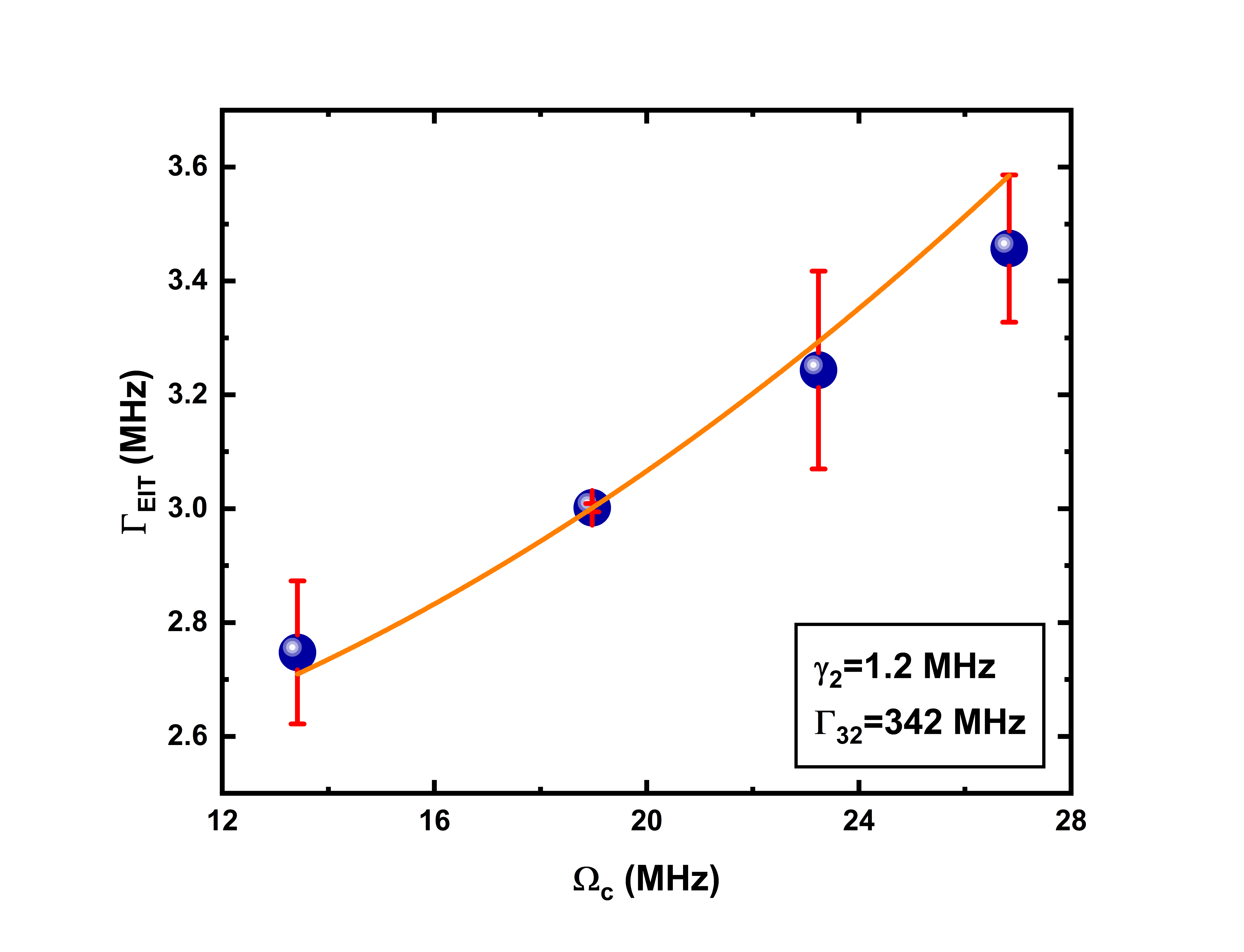}}
         
         \label{three}
     \end{subfigure}
     \hfill
     \begin{subfigure}[b]{0.45\textwidth}
         \centering
         \fbox{\includegraphics[width=\textwidth, height=4.5cm]{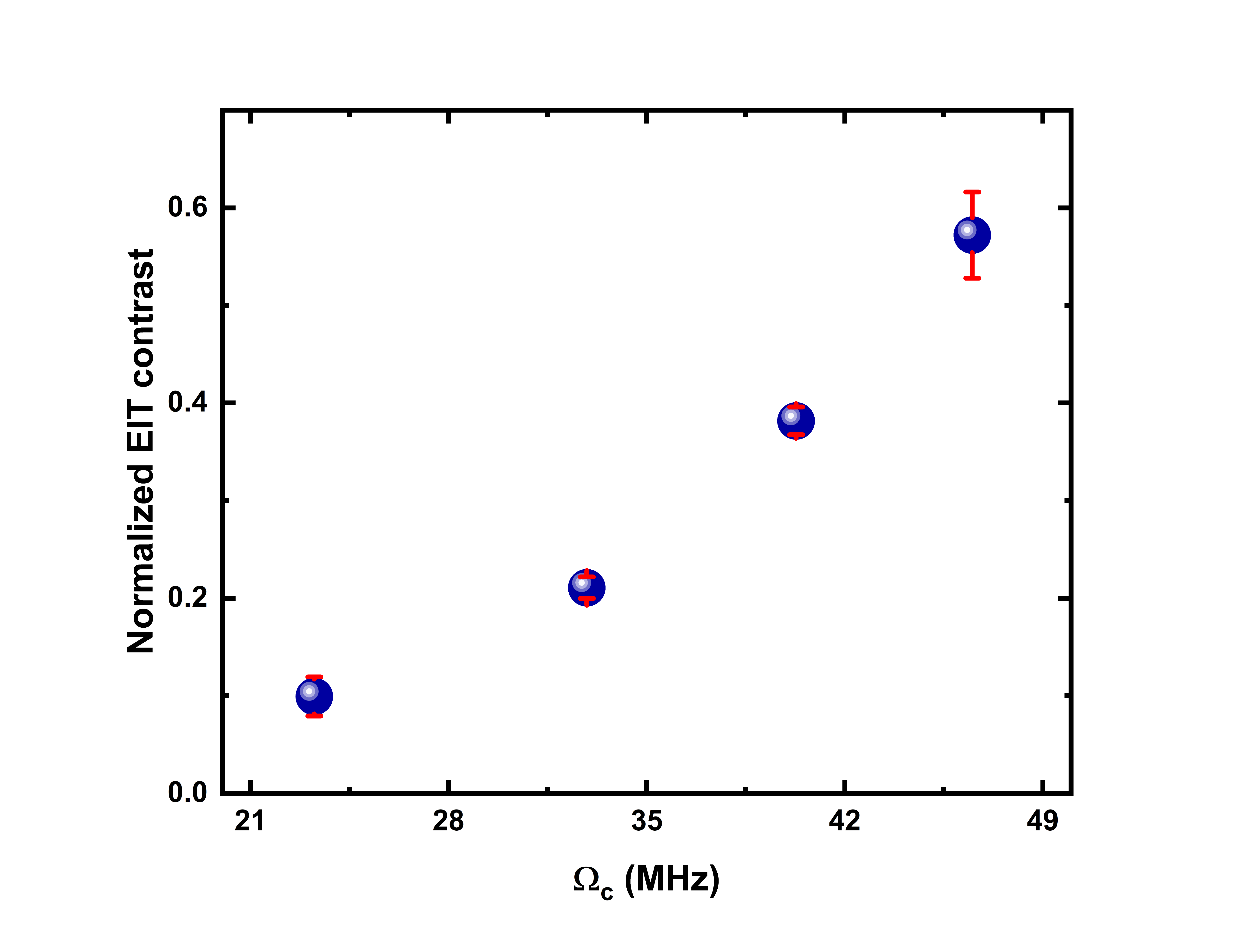}}
        
         \label{on}
     \end{subfigure}
        \caption{(Left) EIT line width is plotted against control Rabi frequency $\Omega_c$ with error bars. The blue-filled circles are experimental data points and the orange solid line is the theoretical fit. (Right) EIT contrast is plotted against $\Omega_c$ with error bars. (color online)}
        \label{sf2}
\end{figure}

\section{EIT line width characteristics}
At $\Delta_c=-923.4$, EIT features are recorded for case $C_1$ for various values of control power. The Gaussian background is subtracted and a Lorentzian fit is performed. Figure \ref{sf2} shows the line width and contrast variation with control Rabi frequency $\Omega_c$. The contrast increases with $\Omega_c$ at the expense of broader line width. By rearranging the imaginary part of probe susceptibility, we can read off the EIT line width
\begin{equation}
    \Gamma_{EIT} = 2[\gamma_2 + \frac{\Omega_c^2}{\sqrt{OD}\Gamma_{32}}]
\end{equation}
Here OD is the optical density at temperature T. The expression to calculate OD \cite{DeRose2023} is given by

\begin{equation}
    OD = \frac{n\omega_p|d_{13}|L}{\epsilon_0 c\hbar(\Gamma_{31}+\Gamma_{32})}
\end{equation}

In the above equation, $n=2.3 \times 10^{10}$ atoms/cc is the atomic density at cell temperature $T=60^o$C, $\omega_p=391.016$ THz is the probe frequency in $D_2$ line, $|d_{13}|=2.46 \times 10^{-29}$ C-m is the dipole matrix element \cite{Tiecke2011PropertiesOP}, $L=75$ mm is the cell length, $\epsilon_0$, $\hbar$, c are vacuum permittivity, reduced Planck's constant, velocity of light respectively and $\Gamma_{31}$,$\Gamma_{32}$ are Doppler broadened spontaneous decay rates. The density at temperature T is calculated using \cite{Tiecke2011PropertiesOP} 
\begin{equation}
    \log p = 7.4077-\frac{4453}{T}
\end{equation}

\section{Acknowledgments} 
The authors thank Dr. Sanjukta Roy for insightful discussions. The authors also thank S. Majumder and A. Misra for their discussions and assistance. The authors also thank the RRI IT section and workshop.

\bibliographystyle{unsrt}  

\bibliography{references}

\end{document}